\documentstyle[12pt,epsfig]{article}
 \newcommand{\be}{\begin{eqnarray}}
 \newcommand{\ee}{\end{eqnarray}}
 \newcommand{\beq}{\begin{equation}}
 \newcommand{\eeq}{\end{equation}}
 \newcommand{\ba}{\begin{array}{1}}
 \newcommand{\ea}{\end{array}}
 \newcommand{\bb}{\begin{thebibliography}}
 \newcommand{\eb}{\end{thebibliography}}

 \newcommand{\abstitle}[1]{{\small {\bf #1}}}
 \newcommand{\absauthor}[1]{{\small {\bf #1}}}
 \newcommand{\address}[1]{{\it #1}}
 
\begin{document}
 \begin{center}
 \abstitle{Saturation of gluon density and soft $pp$ collisions at LHC }\\
 \vspace{0.6cm}
 \absauthor{G. I. Lykasov, A. A.Grinyuk and V. A. Bednyakov } 
\\ [0.6cm]
 \address{Joint Institute for Nuclear Research -
  Dubna 141980, Moscow region, Russia
}
 \end{center}
 \vspace{0.1cm}
%Version of April 27, 2009\\
 \vspace{0.2cm} 
\begin{center}
{\bf Abstract}
\end{center}
 \vspace{0.1cm}
%{\bf
We calculate the unintegrated  gluon distribution 
at low intrinsic transverse momenta and its parameters
are found from the best description of the SPS and LHC 
data on the $pp$ collision in the soft kinematical 
region. It allows us to study the saturation of the 
gluon density at low $Q^2$  more carefully and find the 
saturation scale. 
%}
\\
%\vspace{0.5cm}
%{\bf PACS:} 12.38 AW, 12.39 Pn -Gluons and quarks,
%13.85 Hd-inelastic scattering: many particle final states

\section{Introduction}
\label{1}
%%%%%%%%%%%%%%%%%%%%%%%%%%%%%%%%%%%%%%%%%%%%%%%%%%%%%%%%%%%%%%%%%%%%%%%%%%%%%%%%%%%%%%%%%
%%%%%%%%%%%%%%%%%%%%%%%%%%%%%%%%%%%%%%%%%%%%%%%%%%%%%%%%%%%%%%%%%%%%%%%%%%%%%%%%%%%%%%%%%%%%%%%
As is well known, hard processes involving incoming protons, such as deep-inelastic 
lepton-proton
scattering (DIS), are described using the scale-dependent parton density functions. 
%(PDFs).
Usually, these quantities are 
calculated as a function of the Bjorken variable $x$
and the square of the four-momentum transfer $q^2=-Q^2$
within the framework of popular collinear QCD factorization based on the DGLAP evolution equations
\cite{{DGLAP}}.
However, for semi-inclusive processes (such as inclusive jet production in DIS, 
electroweak boson production \cite{Ryskin:2003,Ryskin:2010}, etc.) at high energies it is more appropriate
to use the parton distributions unintegrated over the transverse momentum $k_t$ in the framework 
of $k_t$-factorization QCD approach \cite{kT}, see, for example, reviews \cite{Andersson:02,kTreview} 
for more information. 
The $k_t$-factorization formalism is based on the BFKL \cite{BFKL} or 
CCFM \cite{CCFM} evolution equations and 
provides solid theoretical grounds for the effects of initial gluon radiation and intrinsic
parton transverse momentum $k_t$.
The theoretical analysis of the unintegrated quark $q(x,k_t)$ distribution (u.q.d.) and gluon 
$g(x,k_t)$ distribution (u.g.d.) can 
be found, for example, in \cite{GBW:98}-\cite{Nikol_Zakhar:91}.
In this paper 
we estimate the u.g.d. at low intrinsic transverse
momenta $k_t\leq 1.5-1.6$ GeV$/$c and its parameters extracted from the best description of the
LHC data at low transverse momenta $p_t$ of the produced hadrons. We also show that our u.g.d.
similar to the u.g.d. obtained in \cite{GBW:98,Jung:04} at large $k_t$
and different from it at low $k_t$. 
%%%%%%%%%%%%%%%%%%%%%%%%%%%%%%%%%%%%%%%%%%%%%%%%%%%%%%%%%%%%%%%%%%%%%%%%%%%%%%%%%%%%%%%%%%%%%%%%%%%%%%%%%
%{\bf
The u.g.d. is directly related to the dipole-nucleon cross section within the model proposed in
\cite{GBW:98}, see also \cite{Nikol_Zakhar:91}-\cite{BZK:2004}, that is saturated at low $Q$
or large transverse distances $r\sim 1/Q$ between quark $q$ and antiquark ${\bar q}$ in the 
$q{\bar q}$ dipole created from the splitting of the virtual photon $\gamma^*$ in the $ep$ DIS.
So, we find also a new parameterization for this dipole-nucleon cross section, as a function of
$r$, from the {\it modified} u.g.d. and analyze the saturation effect for the gluon density.  
%We apply also the {\it modified} u.g.d. to describe be the HERA data on the structure functions,
%the longitudinal one $F_L$, the charm $F_{2c}$ and bottom $F_{2b}$ structure functions.
%}
\section{Inclusive spectra of hadrons in $pp$ collisions}
\label{sec:1}
%and \cite{RefJ}
%%%%%%%%%%%%%%%%%%%%%%%%%%%%%%%%%%%%%%%%%%%%%%%%%%%%%%%%%%%%%%%%%%%%%%%%%%%%%%%%%%%%%%%%%%%%%
\subsection{Unintegrated gluon distributions}
\noindent
As was mentioned above, the unintegrated gluon density in a proton 
are a subject of intensive studies, and various approaches to 
investigate these quantities have been proposed \cite{BFKL}.
At asymptotically large energies (or very small $x$) the theoretically correct 
description is given by the BFKL evolution equation \cite{BFKL} where the leading $\ln(1/x)$ contributions are 
taken into account in all orders. 
Another approach, valid for both small and large $x$, is given by the CCFM  
gluon evolution equation \cite{CCFM}. It introduces angular ordering of emissions to correctly treat the 
gluon coherence effects. In the limit of asymptotic high energies, it almost equivalent to BFKL \cite{BFKL}, 
but also similar to the DGLAP evolution for large $x \sim 1$. The resulting unintegrated
gluon distribution depends on two scales, the additional scale $\bar q$ is a variable 
related to the maximum angle allowed in the emission and plays the role of the evolution 
scale $\mu$ in the collinear parton densities.
Also it is possible to obtain the two-scale involved unintegrated quark and gluon densities
from the conventional ones using the Kimber-Martin-Ryskin (KMR) prescription \cite{Andersson:02,kTreview}. In
this way the $k_T$ dependence in the unintegrated parton distributions enters only in last step
of the evolution, and usual DGLAP evolution equations can be used up to this step. Such
procedure is expected to include the main part of the collinear higher-order QCD corrections.
Finally, a simple parameterization of the unintegrated gluon density was obtained 
within the color-dipole approach in \cite{GBW:98} on the assumption 
of a saturation of the gluon density at low $Q^2$ which successfully 
described both inclusive and diffracting $EPA$ scattering. This
gluon density $xg(x,k_t^2, Q_0^2)$ is given by
\cite{GBW:98,Jung:04}
\begin{eqnarray}
xg(x,k_t,Q_0)=
\frac{3\sigma_0}{4\pi^2\alpha_s(Q_0)}R_0^2k_t^2
\exp\left(-R_0^2(x)k^2_t\right); 
%\nonumber \\
~R_0=\frac{1}{Q_0}\left(\frac{x}{x_0}\right)^{\lambda/2},
\label{def:GBWgl}
\end{eqnarray}
where $\sigma_0 = 29.12$~mb, $\alpha_s = 0.2$, $Q_0 = 1$~GeV, $\lambda = 0.277$ and
$x_0 = 4.1 \cdot 10^{-5}$.   
This simple expression corresponds 
to the Gaussian form for the effective dipole cross section ${\hat\sigma}(x,r)$
as a function of $x$ and the relative transverse separation ${\bf r}$ of the $q{\bar q}$ pair \cite{GBW:98}. 
In fact, this form could be more complicated. In this paper we study this point and try to find the 
parameterization   
for $xg(x,k_t,Q_0)$, which is related to ${\hat\sigma}(x,r)$, from the best description of the 
inclusive spectra 
of charge hadrons produced in $pp$ collisions at LHC energies and mid-rapidity region.      
%%%%%%%%%%%%%%%%
\subsection{Quark-gluon string model (QGSM) including gluons}
As is well known, the soft hadron production in $pp$ collisions at not large transfer 
can be analyzed within the soft QCD models, namely, the quark-gluon string model (QGSM)
\cite{kaid1}-\cite{BLL:2010} or the dual parton model (DPM) \cite{capell2}. The cut n-pomeron 
graphs calculated within these models result in a reasonable contribution at small but 
nonzero rapidities. 
However, it has been shown recently \cite{BGLP:2011,BGLP:2012} that there are some 
difficulties in using the QGSM to analyze the inclusive spectra in $pp$ collisions 
in the mid-rapidity region and at the initial energies above the ISR one. 
However, it is due to the 
according to Abramovsky-Gribov-Kancheli cutting rules (AGK) \cite{AGK} at mid-rapidity 
($y\simeq 0$), when only one-pomeron Mueller-Kancheli diagrams contribute to the inclusive spectrum 
$\rho_h(y\simeq 0, p_t)$. 
To overcome these difficulties it was assumed  in \cite{BGLP:2011} that there are soft gluons
or the so called {\it intrinsic} gluons in the proton \cite{Brodsky:1981}, which are split  
into $q{\bar q}$ pairs and should vanish at the zero intrinsic transverse momentum ($k_t\sim 0$).
 The total spectrum $\rho_h(y\simeq 0, p_t)$ was split  into
two parts, the quark contribution $\rho_q(y\simeq 0, p_t)$ and the gluon one 
%\cite{BGLP:2011}
and their energy dependence was calculated  \cite{BGLP:2011,BGLP:2012}
\begin{eqnarray}
\rho(p_t)=\rho_q(x=0,p_t)+\rho_g(x=0,p_t)= 
%\nonumber \\
g(s/s_0)^{\Delta}{\bar\phi}_{q}(0, p_t)+
\left(g(s/s_0^{\Delta}- \sigma_{nd}\right)
{\bar\phi}_g(0, p_t)~.
\label{def:rhoagk}
\end{eqnarray}
Here
\be
\rho_q(x=0,p_t)=g(s/s_0)^{\Delta}{\bar\phi}_{q}(0, p_t)
\label{def:rhoq}
\ee
and
\be
\rho_g(x=0,p_t)=\left(g(s/s_0^{\Delta}- \sigma_{nd}\right)
{\bar\phi}_{g}(0, p_t)
\label{def:rhog}
\ee
%%%%%%%%%%%%%%%%%%%%%%%%%%%%%%%%%%%%%%%%%%%%%%%%%%%%%%%%%%%%%%%%%%%%%%%%%%%%%%%%%%%%
\begin{figure}[h!!]
 %\rotatebox{270}
%\centerline{\includegraphics[width=0.4\textwidth]{rho_2360_s_29july.eps}}
%\centerline{\includegraphics[width=0.6\textwidth]{rho_7000_s_29july2.pdf}}
\centerline{\includegraphics[width=0.6\textwidth]{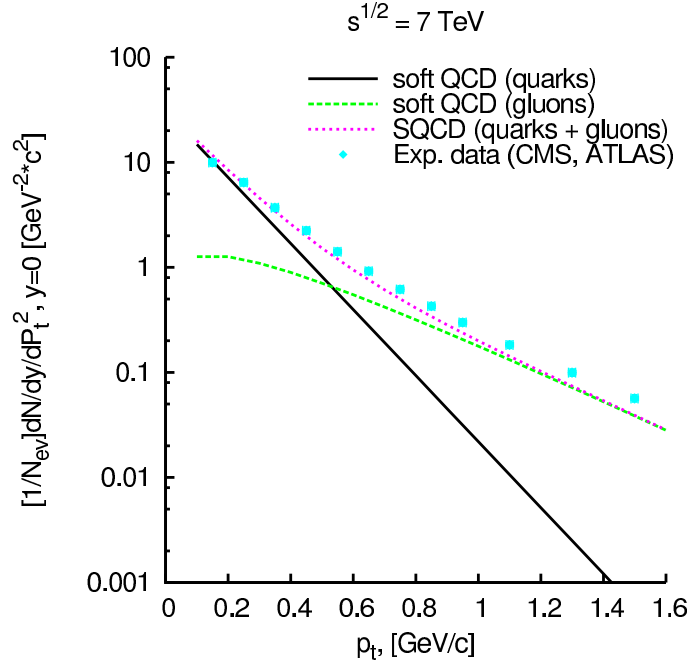}}
%{\epsfig{file=rho_2360_s_29.07.11.eps,height=8cm,width=8cm  }}
%{\epsfig{file=rho_7000_s_29.07.11.eps,height=8cm,width=8cm  }}
%  {\epsfig{file=7TeVM.eps,height=10cm,width=10cm  }}
% {\epsfig{fileAndrei_G.eps=K53p.eps,height=9cm,width=12cm  }}
  \caption[Fig.2]{The inclusive spectrum of the charged hadrons as a function of $p_t$ (GeV$/c$)
in the central rapidity region ($y=0$) at $\sqrt{s}=$7 TeV 
%and wide region of $p_t$ (top) and the same spectrum 
at $p_t\leq$ 1.6 GeV$/$c compared with the CMS \cite{CMS} which are very close to the
ATLAS data \cite{ATLAS}. The solid line is the quark contribution $\rho_q(x=0,p_t)$
(Eq.(\ref{def:rhoq}), the long dashed curve corresponds to the gluon one $\rho_g(x=0,p_t)$ (Eq.(\ref{def:rhog}), 
the dotted line is the sum of the quark and gluon contributions (Eq.(\ref{def:rhoagk}).  
}
\label{Fig_1}
\end{figure} 
%%%%%%%%%%%%%%%%%%%%%%%%%%%                     %%%%%%%%%%%%%%%%%%%%%%%%%%%%%%%%%%%%
\begin{figure}[h!]
\begin{tabular}{cc}
\hspace{1cm}%
\mbox{\epsfig{file=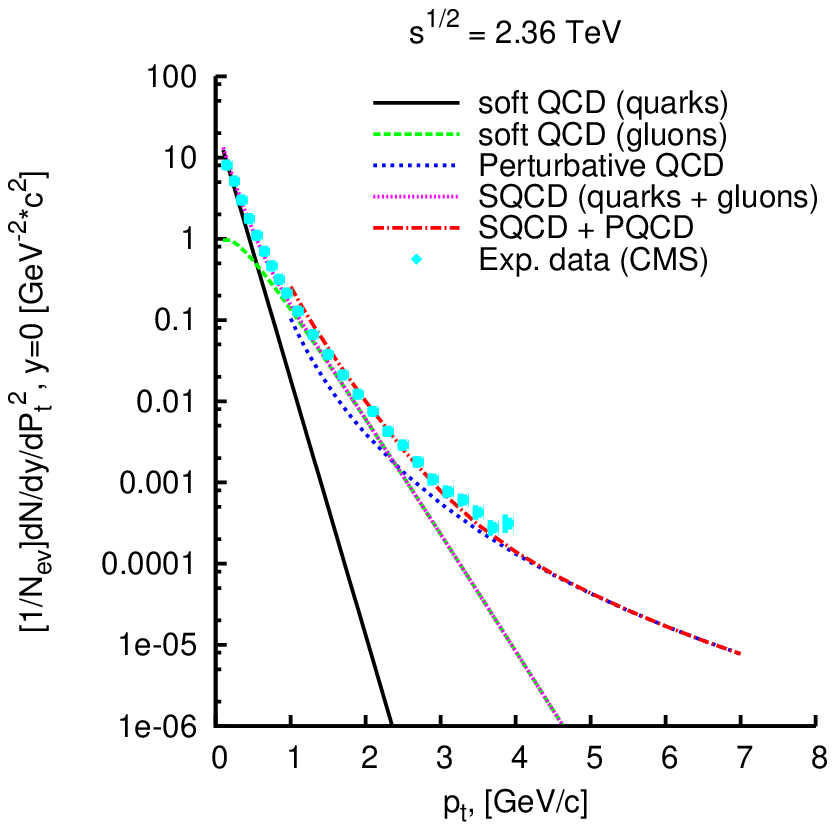,width=0.45\linewidth}}&
\mbox{\epsfig{file=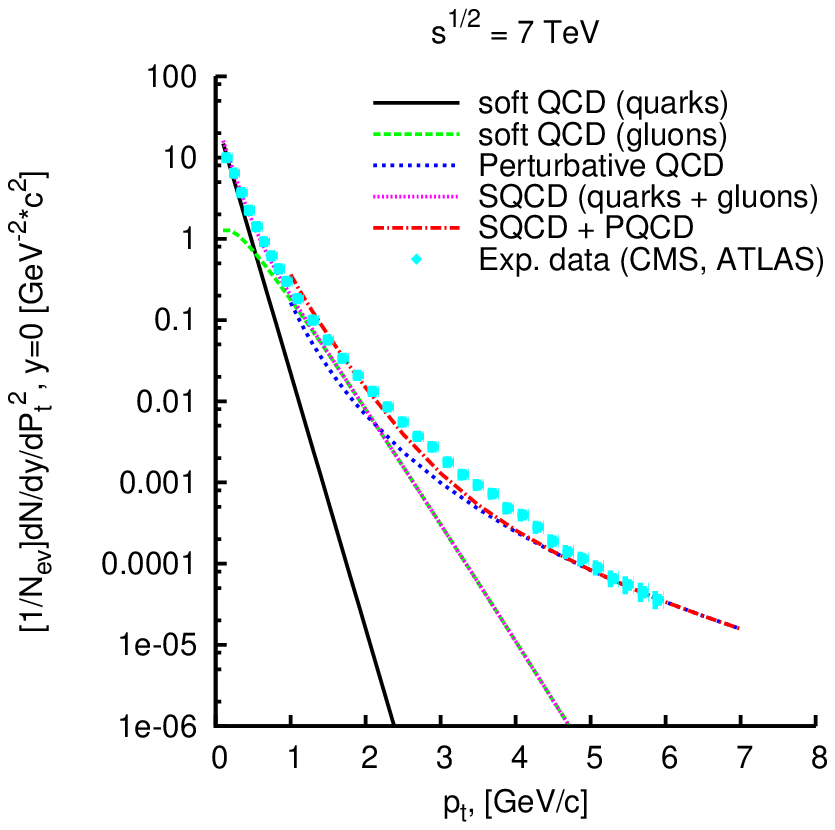,width=0.45\linewidth}}
%\mbox{\epsfig{file=dCSdPt2_10TeV.eps,width=0.43\linewidth}}%
\end{tabular}
%\end{center}
 \caption{The inclusive spectrum of charged hadron as a function of $p_t$ (GeV$/c$)
in the central rapidity region ($y=0$) at $\sqrt{s}=$2.36 TeV (left) and $\sqrt{s}=$7 TeV (right)
 compared with the CMS \cite{CMS} and ATLAS \cite{ATLAS} data. The solid line (soft QCD (quarks))
 is the quark contribution $\rho_q(x=0,p_t)$
(Eq.(\ref{def:rhoq})), the dotted curve (soft QCD (gluons))
corresponds to the gluon one $\rho_g(x=0,p_t)$ (Eq.(\ref{def:rhog}), the long dashed 
line (SQCD(quarks+gluons))
is the sum of the quark and gluon contributions (Eq.(\ref{def:rhoagk}), the short dashed curve (Perturbative QCD)
 corresponds to the perturbative
LO QCD \cite{BGLP:2012} and the dash-dotted line (SQCD+PQCD) is the sum of the calculations within the soft QCD 
including the gluon 
contribution (Eq.(\ref{def:rhoagk})) and the perturbative LO QCD.
} 
\label{Fig_2}
\end{figure}
\begin{figure}[h!!]
\begin{tabular}{cc}
\hspace{1cm}%
\mbox{\epsfig{file=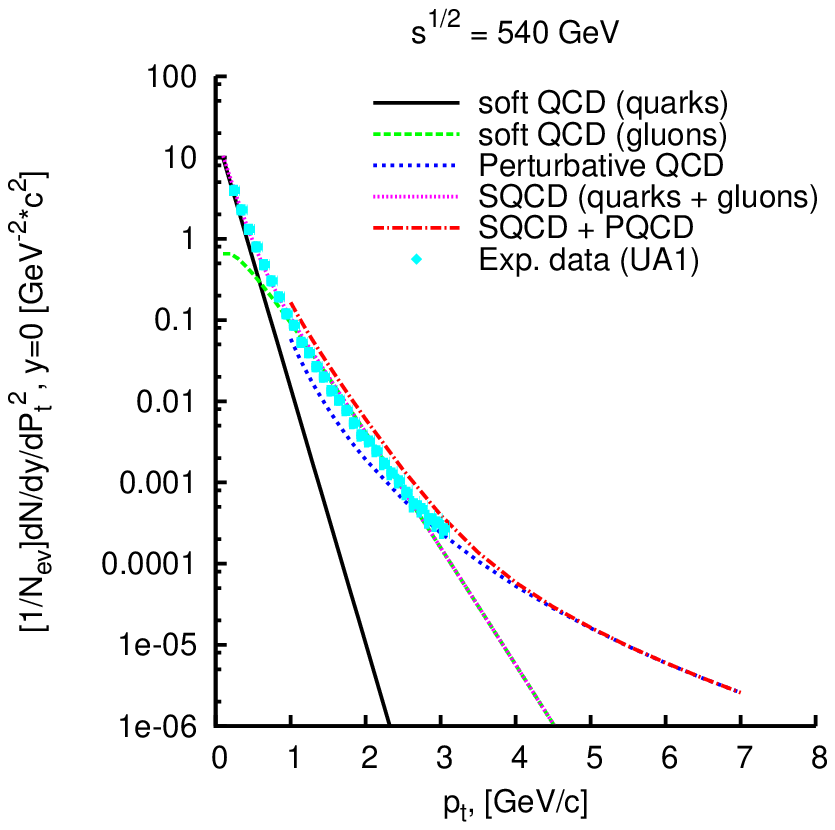,width=0.45\linewidth}}&
\mbox{\epsfig{file=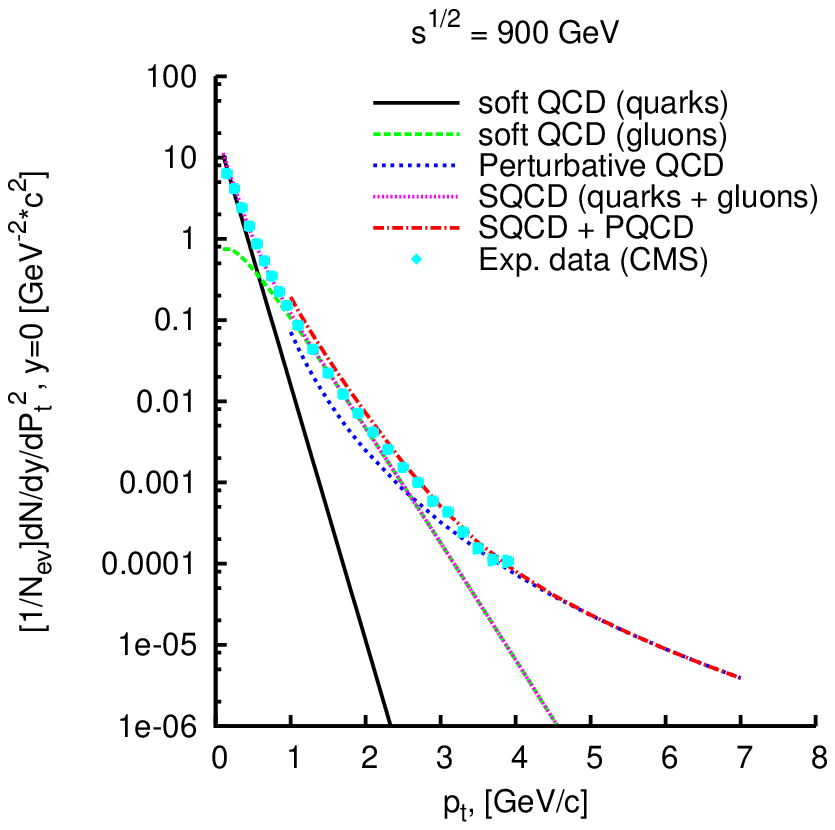,width=0.45\linewidth}}
%\mbox{\epsfig{file=dCSdPt2_10TeV.eps,width=0.43\linewidth}}%
\end{tabular}
%\end{center}
 \caption{The inclusive spectrum of charged hadron as a function of $p_t$ (GeV$/c$)
in the central rapidity region ($y=0$) at $\sqrt{s}=$540 GeV (left) and $\sqrt{s}=$900 GeV (right)
 compared with the UA1 \cite{UA1} and ATLAS \cite{ATLAS} data. The solid line (soft QCD (quarks))
is the quark contribution $\rho_q(x=0,p_t)$
(Eq.(\ref{def:rhoq}), the dotted curve (soft QCD (gluons)) corresponds to the gluon one 
$\rho_g(x=0,p_t)$ (Eq.(\ref{def:rhog}), the long 
dashed line (SQCD(quarks+gluons)) 
is the sum of the quark and gluon contributions (Eq.(\ref{def:rhoagk}), the short dashed curve (Perturbative QCD) 
corresponds to the perturbative LO QCD \cite{BGLP:2012} and the dash-dotted line (SQCD+PQCD)
is the sum of the calculations within 
the soft QCD including the gluon contribution (Eq.(\ref{def:rhoagk})) and the perturbative LO QCD.
} 
\label{Fig_3}
\end{figure}
%%%%%%%%%%%%%%%%%%%%%%%%%%%%%%%%%%%%%%%%%%%%%%%%%%%%%%%%%%%%%%%%%%%%%%%%%%%%%%%%%%%%
The following parameterization for ${\tilde\phi}_q(0,p_t)$ and
 ${\tilde\phi}_g(0,p_t)$ was found \cite{BGLP:2011}:   
\begin{eqnarray}
{\tilde\phi}_q(0,p_t)=A_q\exp(-b_q p_t)~
\nonumber \\
{\tilde\phi}_g(0,p_t)=A_g\sqrt{p_t}\exp(-b_g p_t),
\label{def:phiq}
\end{eqnarray}
where $s_0=1 GeV^2, g=21 mb, \Delta=0.12$.
The parameters are fixed  from the fit to the data on the $p_t$ distribution of
charged particles at $y=0$ \cite{BGLP:2011}:
$A_q=4.78\pm 0.16$ (GeV$/$c)$^{-2}$,~$b_q=7.24\pm 0.11$ (Gev/c)$^{-1}$ and
 $A_g=1.42\pm 0.05$ (GeV$/$c)$^{-2}$;~ 
$b_g=3.46\pm 0.02$ (GeV/c)$^{-1}$. 
Figure 1 illustrates the best fit of the inclusive spectrum of charged hadrons produced in 
$pp$ collisions at $\sqrt{s}=$7 TeV and the central  rapidity region at the hadron transverse momenta
$p_t\leq$ 1.6 Gev$/$c; the solid line corresponds to the quark contribution $\rho_q$,  
the dashed line is the gluon contribution $\rho_g$, and the dotted curve is the sum of these contributions
$\rho_h$ given by Eq.(\ref{def:rhoagk}).

In Figs.~(\ref{Fig_2},\ref{Fig_3}) the inclusive spectra of the charged hadrons produced in $pp$ collisions at the
mid-rapidity region and $\sqrt{s}=$2.36 GeV, 7 Tev, 500 GeV, 900 GeV are presented with the inclusion of the calculations
within the perturbative LO QCD, see details in \cite{BGLP:2012}. One can see from Figs.~(\ref{Fig_1}-\ref{Fig_3}) that
the calculations within the soft QCD including both the quark contribution (Eq.(\ref{def:rhoq})) and the gluon one
(Eq.(\ref{def:rhog})) and the contribution, which corresponds to the LO QCD calculation, results in the satisfactorily
description of these spectra in the wight region of the initial energies. It stimulates us to study the form
of the gluon contribution ${\bar\phi}_g(x~0, p_{ht})$ related to $\rho_g(x~0,p_t)$ in detail and find the information on
the distribution of the soft gluons at small transverse momenta.    
%%%%%%%%%%%%%%%%%%%%%%%%
\subsection{Modified unintegrated gluon distributions}
\begin{figure}[h!]
%\begin{center}
\centerline{\includegraphics[width=0.6\textwidth]{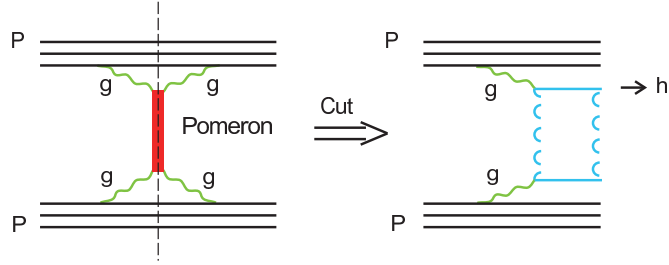}}
% {\epsfig{Pomeron_gl.eps,height=3.cm,width=14.cm  }
% {\epsfig{file=Pomeron_gl.eps,width=1.00\linewidth }}
\caption %[Fig.2]
    {The one-pomeron exchange graph between two gluons in the elastic $pp$ scattering (left) 
and the cut one-pomeron due to the creation of two colorless strings between 
quarks /antiquarks (right) \cite{kaid1}.} 
\label{Fig_Pomeron}
%\end{center}
\end{figure}

As can be seen in Figs.~(\ref{Fig_1}-\ref{Fig_3}) the contribution to the inclusive spectrum at $y\simeq 0$ due to
the {\it intrinsic} gluons is sizable at low $p_t<$ 2 GeV$/$c, e.g., in the soft kinematical region. Therefore,
we can estimate this contribution within the nonperturbative QCD model, similar to the QGSM \cite{kaid1}.   
We calculate the gluon contribution  ${\tilde\phi}_g(x\simeq 0,p_t)$ entering into 
Eq.(\ref{def:phiq}) as the cut graph (Fig.~\ref{Fig_Pomeron}, right) of the one-pomeron exchange in 
the gluon-gluon interaction (Fig.~\ref{Fig_Pomeron}, left) 
using the splitting of the gluons into the $q{\bar q}$ pair.  
The right diagram of  Fig.~\ref{Fig_Pomeron} corresponds to the creation of two colorless
strings between the quark/antiquark $(q/{\bar q})$ and antiquark/quark $({\bar q}/q)$. Then,
after their brake $q{\bar q}$ are produced and fragmented to the hadron $h$.   
Actually, the calculation can be  
made in a way similar to the calculation of the sea quark contribution to the inclusive 
spectrum within the QGSM \cite{kaid1}, 
e.g., the contribution  ${\tilde\phi}_g(0,p_t)$ is presented as the sum of the product of
two convolution functions
\begin{eqnarray}
{\tilde\phi}_g(x,p_t)=F_q(x_+,p_{ht})F_{\bar q}(x_-,p_{ht})+F_{\bar q}(x_+,p_{ht})F_q(x_-,p_{ht})~,
\label{def:tphiq}
\end{eqnarray}
where the function $F_{q({\bar q})}(x_+,p_{ht})$ corresponds to the production of final hadrons
from decay of $q{\bar q}$ string. It is calculated as the following convolution:
\begin{eqnarray}
F_{q({\bar q})}(x_\pm,p_t;p_{ht})=
%\nonumber \\
\int_{x\pm}^1dx_1
\int d^2k_{1t}f_{q({\bar q})}(x_1,k_{1t})G_{q({\bar q})\rightarrow h}
\left(\frac{x_\pm}{x_1},p_{ht}-k_t)\right)~,
\label{def:Fqbrq}
\end{eqnarray}
Here $G_{q({\bar q})\rightarrow h}(z,{\tilde k}_t)=zD_{q({\bar q})\rightarrow h}(z,{\tilde k}_t)$,
$D_{q({\bar q})\rightarrow h}(z,{\tilde k}_t)$ is the fragmentation function (FF) of the quark (antiquark)
to a hadron $h$, $z=x_\pm/x_1,{\tilde k}_t=p_{ht}-k_t$, 
$x_{\pm}=0.5(\sqrt{x^2+x_t^2}\pm x), x_t=2\sqrt{(m_h^2+p_t^2)/s}$.  
The distribution of sea quarks (antiquark) 
$f_{q({\bar q})}$ is related to the splitting function ${\cal P}_{g\rightarrow q{\bar q}}$ of gluons to 
$q{\bar q}$ by
 \begin{eqnarray}
f_{q({\bar q})}(z,k_t)=\int_z^1 g(z_1,k_t,Q_0){\cal P}_{g\rightarrow q{\bar q}}
(\frac{z}{z_1})\frac{dz_1}{z_1}~,
\label{def:fqbq}
\end{eqnarray}
where $g(z_1,k_{1t},Q_0)$ is the u.g.d.. The gluon splitting function ${\cal P}_{g\rightarrow q{\bar q}}$
was calculated within the Born approximation.
In Eq.(\ref{def:fqbq}) we assumed the collinear splitting of the {\it intrinsic} gluon to the $q{\bar q}$ pair
because values of $k_t$ are not zero but small. 

Calculating the diagram of Fig.~(\ref{Fig_Pomeron}) (right) by the use of Eqs.(\ref{def:rhog}-\ref{def:fqbq}) for the gluon 
contribution $\rho_g$ we took the FF to charged hadrons, pions, kaons, and $p{\bar p}$ pairs obtained
within the QGSM \cite{Shabelsk:1992}. From the best description of $\rho_g(x\simeq 0,p_{ht}$, see
its parameterization given by Eq.(\ref{def:phiq}), we found the form     
for the $xg(x,k_t,Q_0)$ which was fitted in the following form:
\begin{eqnarray}
xg(x,k_t,Q_0)=\frac{3\sigma_0}{4\pi^2\alpha_s(Q_0)} 
C_1 (1-x)^{b_g}\times
\nonumber \\
\left(R_0^2(x)k_t^2+C_2(R_0(x)k_t)^a\right)
\exp\left(-R_0(x)k_t-d(R_0(x)k_t)^3\right)~,
\label{def:gldistrnew}
\end{eqnarray}
The coefficient $C_1$ was found from the following normalization:
\begin{eqnarray}
g(x,Q_0^2)=\int_0^{Q_0^2} dk_t^2g(x,k_t^2,Q_0^2)~,
\label{def:BFKL}
\end{eqnarray}
and the parameters 
$$
a=0.7; C_2\simeq 2.3; \lambda=0.22; b_g=12; d=0.2; C_3=0.3295
$$
were found from the best fit of the LHC and SPS data on the inclusive spectrum of charged
hadrons produced in $pp$ collisions and in the mid-rapidity region at $p_t\leq$1.6 GeV$/$c,
see the dashed lines (SQCD (quark+gluons)) in Figs.~(\ref{Fig_1}-\ref{Fig_3}) and Eq.(\ref{def:phiq}).

%%%%%%%%%%%%%%%%%%%%%%%%%%%%%%%%%%%%%%%%%%%%%%%%%%%%%%%%%%%%%%%%%%
\begin{figure}[h!]
\centerline{\includegraphics[width=0.6\textwidth]{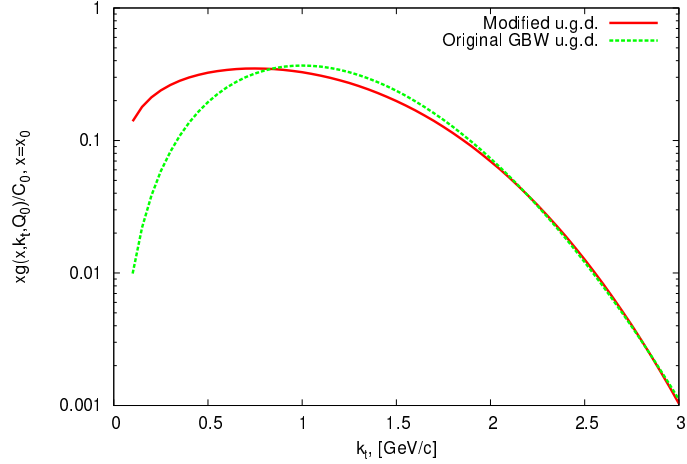}}
%\begin{center}
%%\rotatebox{270}
%{\epsfig{ugdf.ps,height=3.cm,width=14.cm  }}
% {\epsfig{file=ugdf.ps,width=0.60\linewidth }}
\caption %[Fig.2]
    {The unintegrated  gluon distribution $xg(x,k_t,Q_0)/C_0$ as a function of $k_t$
at $x=x_0$ and $Q_0=1.$GeV$/$c. The dashed curve corresponds to the original GBW 
\cite{GBW:98,Jung:04}, Eq.(\ref{def:GBWgl}), and the solid line is the modified u.g.d. 
given by Eq.(\ref{def:gldistrnew}).
% which is very close to the original GBW at 
%$k_t\geq 1.5$ GeV$/$c.
} 
\label{Fig_ugd}
%\end{center}
\end{figure} 
Figure~\ref{Fig_ugd} presents the modified u.g.d. obtained by calculating the cut one-pomeron graph of 
Fig.~\ref{Fig_Pomeron}
%\end{center}
and the original GBW u.g.d. \cite{GBW:98,Jung:04} as a function of the transverse gluon momentum
$k_t$. Here $C_0=3\sigma_0/(4\pi^2\alpha_s(Q_0))$. One can see that the modified u.g.d.
(the solid line in Fig.~\ref{Fig_ugd})
%\end{center}) 
is different from the original GBW u.g.d. \cite{GBW:98,Jung:04} at $k_t~<~1.5$ GeV$/$c
and coincides with it at larger $k_t$. This is due to the sizable contribution of $\rho_g$ 
(Eqs.(\ref{def:rhog},\ref{def:phiq})) to the inclusive spectrum $\rho(p_t)$ of charged hadrons produced
in $pp$ collisions at the LHC and SPS energies in the mid-rapidity region, see the dashed lines (soft QCD(gluons)) in 
Figs.~(\ref{Fig_1}-\ref{Fig_3}). 

Let us also note that, as is shown recently in \cite{GJLLZ:2012}, the modified GBW given by 
Eq.(\ref{def:gldistrnew}) does not contradict the HERA data on the longitudinal structure function 
$F_l(Q^2)$ at low $x$, the charm structure function $F_{2c}(x, Q^2)$ and the bottom one $F_{2b}(x, Q^2)$.
%%%%%%%%%%%%%%%%
\section{Saturation dynamics}
%{\bf
According to \cite{GBW:98}, the u.g.d. can be related to the cross section ${\hat\sigma}(x,r)$ of the 
$q{\bar q}$ dipole with the nucleon. This dipole is created from the split of the virtual exchanged photon
$\gamma^*$ to $q{\bar q}$ pair in the $e-p$ deep inelastic scattering (DIS).This relation at the fixed 
$Q_0^2$ is the following: 
\cite{GBW:98}:
\be
{\hat\sigma}(x,r)~=~\frac{4\pi\alpha_s(Q_0^2)}{3}\int\frac{d^2k_t}{k_t^2}
\left\{ 1-J_0(rk_t)\right\}xg(x,k_t)
\label{def:sigxr}
\ee   
Inserting the simple form for $xg(x,k_t)$ given by Eq.(\ref{def:GBWgl}) to Eq.(\ref{def:sigxr}) 
on can get the following form for the dipole cross section:
\be
{\hat\sigma}_{GBW}(x,r)=\sigma_0\left\{1-exp\left(-\frac{r^2}{4R_0^2(x)}\right)\right\}
\label{def:sigGBW}
\ee
However, the modified u.g.d. given by Eq.(\ref{def:gldistrnew}), inserted to Eq.(\ref{def:sigxr})
results in the more complicated form for ${\hat\sigma}(x,r)$:
\be
{\hat\sigma}_{Modif}(x,r)=\sigma_0\left\{1-exp\left(-\frac{b_1r}{R_0(x)}-\frac{b_2r^2}{R_0^2(x)}
\right)\right\}~,
\label{def:sigxrmod}
\ee
where $b_1=0.045, b_2=0.3$.

\begin{figure}[h!]
\centerline{\includegraphics[width=0.8\textwidth]{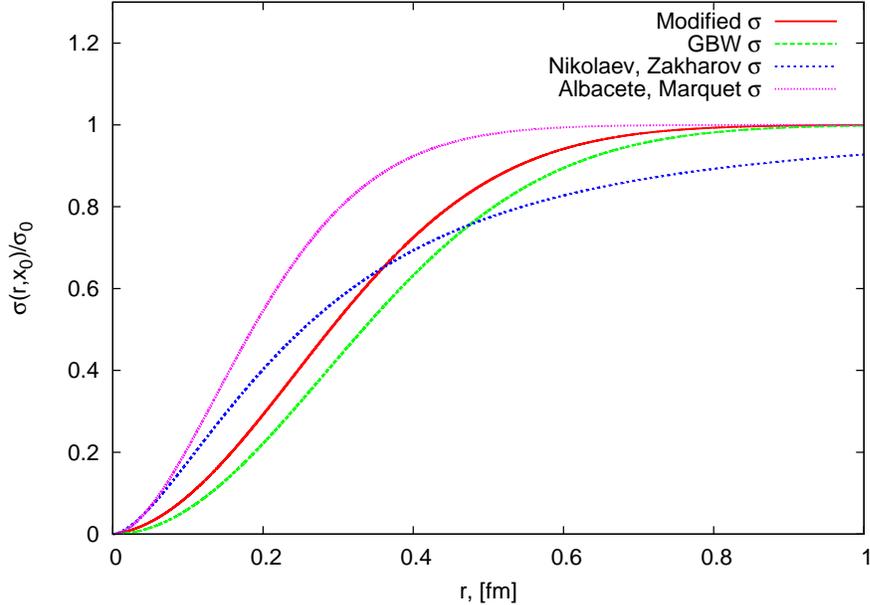}}
%\begin{center}
%%\rotatebox{270}
%{\epsfig{ugdf.ps,height=3.cm,width=14.cm  }}
% {\epsfig{file=ugdf.ps,width=0.60\linewidth }}
\caption %[Fig.2]
    {The dipole cross section ${\hat\sigma}/\sigma_0$ at $x=x_0$ as a function of $r$.
The green line ``GBW $\sigma$'' is calculation of \cite{GBW:98}, the red line 
``Modified $\sigma$'' corresponds to our calculation; the curve "Nikolaev,Zakharov $\sigma$''
is the calculation of \cite{Nikol_Zakhar:91}; the line "Albacete,Marquet $\sigma$'' is
the calculation of \cite{Marquet:2010}.  } 
\label{Fig_5}
%\end{center}
\end{figure}

There are a few forms for the dipole cross sections suggested in 
\cite{Nikol_Zakhar:91}-\cite{BZK:2004}. The dipole cross section can be presented in the general form: 
\cite{GBW:98}:
\be
{\hat\sigma}(x,r)=\sigma_0g({\hat r}^2)~,
\label{def:sigdipgen}
\ee
where ${\hat r}=r/(2R_0(x))$. The function $g({\hat r}^2)$ was presented in the form \cite{Nikol_Zakhar:91} 
\be
g({\hat r}^2)={\hat r}^2ln\left(1+\frac{1}{{\hat r}^2}\right)
\label{def:sigdipNZ90}
\ee
or in the form \cite{Marquet:2010}:
\be
g({\hat r}^2)=1-\exp\left\{-{\hat r}^2ln\left(\frac{1}{\Lambda r}+e\right)\right\}
\label{def:sigdipMV98}
\ee 
which both of them are saturated when $r$ grows.
The function $g({\hat r}^2)$ was also presented in the form of type \cite{Nikol_Zakhar:1994,Levin:1998}:
\be
g({\hat r}^2)=ln(1+{\hat r}^2)
\label{def:sigdipGLM98}
\ee
that is not saturated when $r$ increases.
Figure~(\ref{Fig_5}) illustrates the dipole cross sections ${\hat\sigma}/\sigma_0$ at $x=x_0$ which are saturated 
at $r>0.6$ fm, see \cite{Nikol_Zakhar:91,McLerran:1998,Marquet:2010}.
They are compared with our calculations (solid line, Modified $\sigma$) given by Eq.(\ref{def:sigxrmod}).  
The solid line in Fig.~\ref{Fig_5} corresponds to the modified u.g.d. given by Eq.(\ref{def:gldistrnew})
the application of that allowed us to describe satisfactorily the LHC data on inclusive spectra of hadrons produced in
the mid-rapidity region of $pp$ collision at low $p_t$. Therefore, the form of the dipole-nucleon cross 
sections presented in Fig.~\ref{Fig_5} can be verified by the last LHC data on hadron spectra in soft 
kinematical region.

Comparing the solid line (Modified $\sigma$) and dashed curve (GBW $\sigma$) in Fig.~\ref{Fig_5} one can
see that ${\hat\sigma}_{Modif}(x,r)$ given by Eq.(\ref{def:sigxrmod}) saturates faster than  
${\hat\sigma}_{GBW}(x,r)$ given by Eq.(\ref{def:sigGBW}) when the $q{\bar q}$ dipole distance $r$ increases.
If $R_0=1/GeV(x/x_0)^{\lambda/2}$, according to \cite{GBW:98}, then for saturation scale has the form
$Q_s\sim 1/R_0=Q_{s0}(x_0/x)^{\lambda/2}$, where $Q_{s0}=1$ GeV$=$0.2 fm$^{-1}$. The saturation in the dipole 
cross section (Eq.(\ref{def:sigGBW}) sets in when $r\sim 2R_0$ or $Q_{s}\sim (Q_{s0}/2)(x_0/x)^{\lambda/2})$.
Comparing the saturation properties of the Modified $\sigma$ and GBW $\sigma$ presented in Fig.~\ref{Fig_5}
one can get slightly larger value for $Q_{s0}$ in comparison to $Q_{s0} =$ 1 GeV$/$c.
%}
\section{Conclusion} 
%We fitted the experimental data on the inclusive
%spectra of charged particles produced in the central $pp$ collisions at energies larger 
%%than the ISR starting
% from 500 GeV up to 7 TeV 
%with the sum of the quark contribution $\rho_q$  and the gluon
%contribution $\rho_g$ (see Eqs.(\ref{def:rhoagk},\ref{def:phiq})).
% The parameters of this fit do not depend on the initial energy in that energy interval.
%{\bf
Assuming creation of soft {\it intrinsic} gluons in the proton at low transverse momenta $k_t$ and 
calculating the cut one-pomeron graph between two gluons in colliding protons we found
the form for the unintegrated gluon distribution (modified u.g.d) as a function of $x$ 
and $k_t$ at the fixed value of $Q_0^2$.
The parameters of this u.g.d. were found from the best description of the SPS and LHC data on 
the inclusive spectra of the charged hadrons
produced in the mid-rapidity pp collisions at low $p_t$. It was shown that the modified u.g.d.
is different from the original GBW u.g.d. obtained in  \cite{GBW:98,Jung:04}
at $k_t\leq$ 1.6 GeV$/$c and it coincides with the GBW u.g.d. at $k_t>1.6$GeV$/$c.

Using the modified u.g.d. we calculated the $q{\bar q}$ dipole-nucleon cross section ${\hat\sigma}_{Modif}$ 
as a function of the transverse distance $r$ between $q$ and ${\bar q}$ in the dipole and found that it saturates  
faster than the ${\hat\sigma}_{GBW}$ obtained within the GBW dipole model \cite{GBW:98}. It allowed us to
find the saturation scale $Q_s$ for the gluon density that is larger than the one obtained in \cite{GBW:98}.
Moreover, we showed that the satisfactory description of the LHC data using the modified u.g.d.
can verify the form of the dipole-nucleon cross section and the property of the saturation of the gluon 
density at low $Q^2$.
%}
%%%%%%%%%%%%%%%%%%%%%%%%%%%%%%%%%%%%%%%%%%%%%%%%%%%%%%%%%%%%%%%%%%%%%%%%%%%%%%%%%%%%%%%%%%%%%%%%%%%%%

\vspace{1cm}
{\bf Acknowledgments}\\
The authors are grateful to S.P. Baranov, B.I. Ermolaev, H. Jung, A.V.Kotikov, A.V.Lipatov,  M.G. Ryskin, 
Yu.M. Shabelskiy and N.P.Zotov 
for useful discussions and comments.
This research was supported by the 
RFBR grant 11-02-01538-a.

%%%%%%%%%%%%%%%%%%%%%%%%%%%%%%%%%%%%%%%%%%%%%%%%%%%%%%%%%%%%%%%%%%%%%%%%%%%%%%%%%%%%%%%%%%%%%%%%%%%%%%
\begin{footnotesize}

\end{footnotesize}


\begin{thebibliography}{99}
\bibitem{DGLAP}
{\it Gribov V.N., Lipatov L.N.}// Sov.J.Nucl.Phys. 1972. V.15 P. 438;
{\it Altarelli G., Parisi G.}// Nucl.Phys. 1997. B. V.126.P. 298;
{\it Dokshitzer Yu.L.}// Sov.Phys. JETP. 1977. V. 46. P. 641.
\bibitem{Ryskin:2003}
{\it Watt G., Martin A.D., Ryskin M.G.}// Eur.Phys.J.
2003. C. V. 31. P. 73; [arXiv:hep-ph/0306169];
\bibitem{Ryskin:2010}
{\it Martin A.D., Ryskin M.G., Watt G.} Eur.Phys.J. C. 2010 v. 66. P. 163;
arXiv:/0909.5529 [hep-ph].
\bibitem{kT} 
  {\it Gribov L.V., Levin E.M., Ryskin M.G.}// Phys. Rep. 1983. V. 100. P. 1;\\
  {\it Levin E.M., Ryskin M.G., Shabelsky Yu.M., Shuvaev A.G.}// Sov. J. Nucl. Phys. 
  1997. V. 53. P. 657;\\
  {\it Catani S., Ciafoloni M., Hautmann F.}//, Nucl. Phys. B. 1991. V. 366. P.135;\\
  {\it Collins J.C., Ellis R.K.}// Nucl. Phys. B. 1991. V. 360.P. 3.
\bibitem{Andersson:02}
{\it Andersson Bo., et.al.}// Eur.Phys.J. C. 2002. V. 25. P. 77.
\bibitem{kTreview} 
%B.~Andersson {\sl et al.} (Small-$x$ Collaboration), Eur. Phys. J. C {\bf 25}, 77 (2002);\\
 {\it Andersen J. {\sl et al.} (Small-$x$ Collaboration)}// Eur. Phys. J. C.
 2004. V. 35. P. 67;
  {\it ibid} Eur. Phys. J. C. 2006. V. 48. P. 53.
\bibitem{BFKL}
{\it Lipatov L.N.}// Sov.J.Nucl.Phys. 1976. V. 23. P. 338;
{\it Kuraev E.A., Lipatov L.N., Fadin V.S.}// Sov.Phys. JETP. 1976. V. 44. P. 443;
{\it ibid}  1977. V. 45. P.199; 
{\it Balitzki Ya.Ya., L.N.~Lipatov L.N.}// Sov.J. Nucl.Phys. 
1978. V. 28. P. 822;
{\it L.N.~Lipatov}// Sov.Phys. JETP. 1986. V. 63. P. 904.
\bibitem{CCFM}
{\it Ciafaloni M.}// Nucl.Phys. B 1988. V. 296. P. 49; 
{\it Catani C., Fiorani F., Marchesini G.}//
Phys.Lett. B. 1990. V. 234. P. 339; {\it ibid} Nucl.Phys. B 
1990. V. 336. P. 18;
{\it Marchesini G.}// Nucl.Phys. B. 1995. V. 445. P. 49; 
{\it ibid} Phys.Rev. D. 2004. V. 70. P. 014012;
[Errartum-ibid. D. 2004. V. 70. P. 079902] [arXiv:hep-ph/0309096].
\bibitem{GBW:98}
{\it Golec-Biernat K., Wusthof M.}//
Phys.Rev. D. 1998. V. 59. P. 014017; 
{\it ibid} Phys.Rev. D. 1999. V. 60. P. 114023.
\bibitem{Jung:04}
{\it Jung H.}// Proc. of the DIS'2004, Strbaske' Pleso, Slovakia,
arXiv:0411287 [hep-ph].
%\bibitem{Ryskin:2010}
%{\it Martin A.D., Ryskin M.G., Watt G.}// Eur.Phys.J. C. 2010. V. 66. P. 163;
%arXiv:/0909.5529 [hep-ph].
\bibitem{KLZ:02}
{\it Kotikov A.V., Lipatov A.V., Zotov N.P.}// Eur.Phys.J. C. 2003. V. 27. P. 219.
\bibitem{Ermol1:11}
{\it Ermolaev B.I., Greco V., Troyan S.I.}//
Eur.Phys.J. C. 2011. V. 71. P. 1750.
\bibitem{Nikol_Zakhar:91}
{\it Nikolaev N.N., Zakharov B.G.}//
Z.Phys. C. 1991. V. 49. P. 607.
\bibitem{Nikol_Zakhar:1994}
{\it Nikolaev N.N., Predazzi E., Zakharov B.G.}// 
Phys.Lett. B. 1994. V. 326. P. 161.
\bibitem{Levin:1998}
{\it Gostman E., Levin E., Maor U.}// 
Phys.Lett. B. 1998. V. 425. P. 369.
%%%%%%%%%%%%%
\bibitem{McLerran:1998} 
{\it McLerran L.D., Venugopalan R.}// 
%Boost covariant gluon distributions in large nuclei, 
Phys. Lett. B. 1998. V. 424. P. 15; arXiv:9705055 [nucl-th]. 
%doi:10.1016/S0370-2693(98)00214-7.
\bibitem{Marquet:2010}
%Single Inclusive Hadron Production at RHIC and the LHC from the Color Glass Condensate.
{\it Albacete J. L., Marquet C.}// 
Phys.Lett. B. 2010. V. 687. P. 174; arXiv:1001.1378 [hep-ph].
\bibitem{BZK:2004}
{\it Kopeliovich B.Z., Raufeisen J.}// 
Lect. Notes Phys. 2004. V. 647 P. 305.
%%%%%%%%%%%%%%%%   %%%%%%%%%%%%%%%%%%%%%%%%%     %%%%%%%%%%%%%%%%%%%%%%%%%%%%%%%%%%%%
\bibitem{kaid1}
{\it Kaidalov A.B.}//
Z.Phys. 1982. C. V. 12. P. 63; Sarveys High Energy Phys. 
1999. V. 13. P. 265;
{\it Kaidalov A.B., Piskunova O.I.}// Z.Phys. C. 1986. V. 30 P. 145;
Yad.Fiz. 1986. V. 43. P. 1545.
\bibitem{LS:1992}
{\it Lykasov G.I., Sergeenko M.N.}// Z.Phys. C. 1992. V. 56. 697;  
{\it ibid} Z.Phys. C. 1991. V. 52. P. 635;
{\it ibid} Z.Phys. C. 1996. V. 70. P. 455.
\bibitem{LKSB:2009}
{\it Lykasov G.I., Karpova Z.M., Sergeenko M.N., Bednyakov V.A.}
Europhys.Lett. V. 86. 2009. P. 61001; PoS 2008LHC (2008) 119.
\bibitem{BLL:2010}
 {\it Bednyakov V.A., Lykasov G.I., Lyubushkin V.V.}//
Europhys.Lett. 2010. V. 92. P. 31001;
  arXiv:1005.0559 [hep-ph].
\bibitem{capell2}
{\it Capella A., Sukhatme U., Tan C.J., Tran Thanh Van J.}// 
Phys.Rev. D. 1987. V.36. P.109;
{\it ibid} Adv.Ser.Direct.High Energy Phys. 1988. V. 2. P. 428.
\bibitem{BGLP:2011}
{\it Bednyakov V.A., Grinyuk A.A., Lykasov G.I., Poghosyan M.}
Nucl.Phys. B {\bf 219-220} [Proc.Suppl.] (2011) 225;
arXiv:11040532 [hep-ph].
\bibitem{BGLP:2012}
{\it Bednyakov V.A., Grinyuk A.A., Lykasov G.I., Poghosyan M. }//
Int.J.Mod.Phys. A. 2012. V. 27. P. 1250042.
\bibitem{AGK}
{\it Abramovsky V., Gribov V.N., Kancheli O.}//
 Sov.J.Nucl.Phys. 1973. V. 18. P. 308.
\bibitem{Brodsky:1981}
{\it Brodsky S., Peterson C., Sakai N.}//
 Phys.Rev. D. 1981. V. 23. P. 2745.
\bibitem{CMS}
{\it Kachatryan Vadran, et al., (CMS Colalboration)}//  
Phys.Rev.Lett. 2010. V, 105. P. 022002; arXiv:1005.3299 [hep-ex]; 
CMS-QCD, CERN-PH-EP-2010-003, Feb.2010.
\bibitem{ATLAS}
{\it  Aad G., et al., (ATLAS Collaboration)}// 
New J.Phys. 2011. V. 13. P. 053033;
arXiv:1012.5104v2 [hep-ex]. 
\bibitem{UA1}
{\it C.Albajar, et al., (UA1 Collaboration)} 
Nucl.Phys. B. 1990. B. V. 335. P. 261.
%\bibitem{Brodsky:1981}
%{\it Brodsky S., Peterson C., Sakai N.}//
% Phys.Rev. D. 1981. V. 23. P. 2745.
\bibitem{Shabelsk:1992}
{\it Shabelsky Yu.M.}// 
Sov.J.Nucl.Phys. 1992. V. 56. P. 2512.
\bibitem{GJLLZ:2012}
{\it Grinyuk A.A., et al.,}// 
Proc. of 3rd International Workshop on Multiple Parton Conference: 
C11-11-21. 2012. PP.169-176; arXiv:1203.0939 [hep-ph]; arXiv:1301.4545 [hep-ph]

%%%%%%%%%%%%%%%    %%%%%%%%%%%%%%%%%%%%%%%%      %%%%%%%%%%%%%%%%%%%%%%%%%%%%%%%%%%%%

\end{thebibliography}
\end{document}